\documentclass[a4paper,twoside,10pt]{article}
\usepackage{graphicx,publaob}
\usepackage{amsmath}
\usepackage{hyperref}
\usepackage{enumitem}

\def\papertype{Invited Lecture}
\begin{document}

\title{X-shaped radio galaxy 3C 315}

\authors{V. Borka Jovanovi\'{c}$^1$, D. Borka$^1$ \lowercase{and} P. Jovanovi\'{c}$^2$}

\address{$^1$Department of Theoretical Physics and Condensed Matter Physics (020),\\
Vin\v{c}a Institute of Nuclear Sciences - National Institute of the Republic of Serbia,\\
University of Belgrade, P.O. Box 522, 11001 Belgrade, Serbia}
\Email{vborka}{vinca}{rs}
\Email{dusborka}{vinca}{rs}
\address{$^2$Astronomical Observatory, Volgina 7, P.O. Box 74, 11060 Belgrade, Serbia}
\Email{pjovanovic}{aob}{rs}

\markboth{X-shaped radio galaxy 3C 315}{V. Borka Jovanovi\'{c} et al.}

\abstract{Here we want to investigate X-shaped radio galaxy 3C 315, which is a FRII source, but it lies very close to the FRI/FRII borderline. We used publicly available data from Leahy’s atlas of double radio-sources and NASA/IPAC Extragalactic Database (NED) in order to investigate its flux density, as well as the spectral index distribution. We obtained spectral index distributions between the frequencies: 1417 MHz - 2695 MHz and 1646 MHz - 2695 MHz. Our conclusion is that the synchrotron radiation is the dominant radiation mechanism over most of the area of 3C 315. Because of very poor hotspots (i.e. warmspots) we investigated which part of the source represents more active regions. The results of this study would be helpful to understand the evolutionary process of the galaxy 3C 315.}

\section{INTRODUCTION}

DRAGNs are Double Radio sources Associated with Galactic Nuclei. They are clouds of radio-emitting plasma which have been shot out of active galactic nuclei (AGN) via narrow jets. As explained by Leahy (1993), a DRAGN is a radio source containing at least one of the following types of extended, synchrotron-emitting structures: jet, lobe, and hotspot complex.

The most of radiation emitted from AGNs is at radio wavelengths. Some of the radio galaxies has X-shaped, and it is called X-shaped or winged radio galaxies. The origin of the X-shaped radio morphology could be explained using the following possibilities:
\begin{enumerate}[label=\arabic*),nosep]
\item the AGN has undergone two separate epochs of activity. The brighter lobes define the axis of current activity (Merritt and Ekers 2002);
\item the radio jets are expanding into an asymmetric medium, causing backflow and producing secondary wings (Leahy and Williams 1984; Capetti 2002);
\item the coalescence of two supermassive black holes (SMBH) previously hosted by a pair of merging galaxies (Gopal-Krishna, Biermann and Wiita 2003).
\end{enumerate}

X-shaped structure characterize pairs of lobes centered on a AGN, where one lobe pair is usually brighter then other (Saripalli et al. 2007).

DRAGN 3C 315 is an X-shaped radio source (it has two sets of double lobes angled with respect to each other). From one hand 3C 315 is a well X-shaped source and example where both lobe pairs have FR-I structure (Saripalli et al. 2007). The host galaxy also has high ellipticity and wings close to the minor axis. This source contains only a weak hotspot (warmspot) in the far-side of the north-western lobe, but on the other hand it would still be classified as an FR-II radio galaxy. It was shown that the host galaxy of 3C 315 (Koff et al. 2000) is highly elongated in the north west - south east direction and accompanied by an elliptical galaxy.

Many authors investigated radio galaxy 3C 315. Northover (1976) presented high-resolution maps of 3C 315 made at Cambridge with One-Mile and 5-km radio telescopes. He find that source consists of two extended components on either side of the central object and that the source has a straight spectrum over a wide frequency range suggesting that electrons are continually being supplied from the central object (Northover 1976).

Hogbom presented observation at 4995 MHz and find that the source 3C 315 is strongly polarized (Hogbom 1979). Leahy and Williams presented maps of 3C 315 at 1.4 GHz (Leahy and Williams 1984). Alexander and Leahy presented intensity images of 3C 315 at 1.4, 1.6 and 2.7 GHz (Alexander and Leahy 1987) and made it publicly available via NED (Leahy, Bridle and Strom 2013). They find that 3C 315 is a very unusual source, both structurally and in its spectral behavior. Leahy, Pooley and Rileyet (1986) also studied polarization of 3C 315. There are also other researchers who investigated radio source 3C 315 (Saripalli et al. 2007; Saripalli and Subrahmanyan 2009; Marecki 2012; Yang 2019).

\section{THE PROTOTYPE OF X-SHAPED (WINGED) DRAGN: 3C 315}

The radio source 3C 315 (cross-identifications: 3C 315; 4C +26.47; PKS 1511+26; B2 1511+26; LQAC 228+026 002) is the prototype X-shaped or winged DRAGN. Wings could be described as a secondary pair of lobes, lying at some angle from the main axis defined by the wormspots and the brighter parts of the diffuse lobes.

Unlike most other cases, there are no true hotspots in either pair of lobes, although there is a modest warmspot about three-quarters of the way to the end of the North North East (NNE) lobe.

\begin{figure*}[ht!]
\centering
\includegraphics[width=0.48\textwidth]{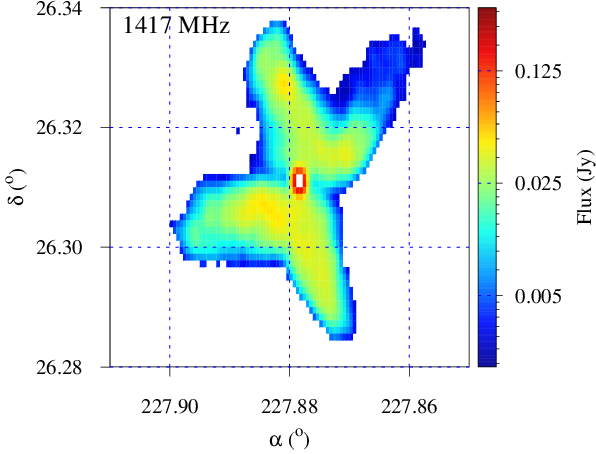}
\hfill
\includegraphics[width=0.46\textwidth]{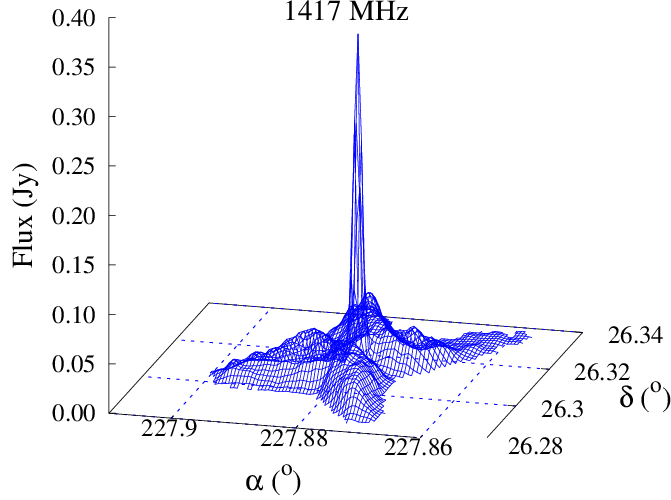}
\caption{2D plot (left) and 3D plot (right) of 3C 315 flux density distribution (in Jy), presented in Equatorial coordinate system $(\alpha, \delta)$, at 1417 MHz.}
\label{fig01}
\end{figure*}

\begin{figure*}[ht!]
\centering
\includegraphics[width=0.48\textwidth]{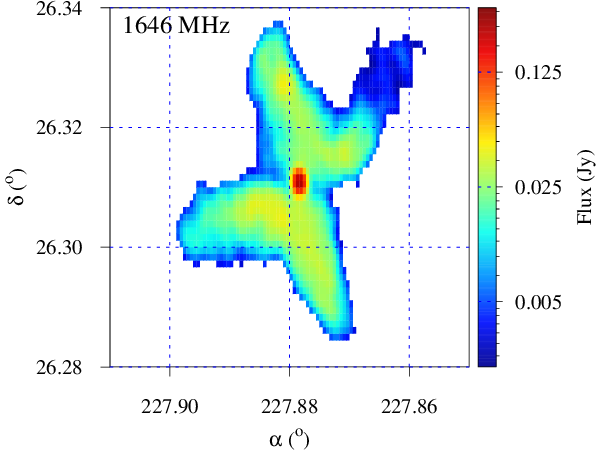}
\hfill
\includegraphics[width=0.46\textwidth]{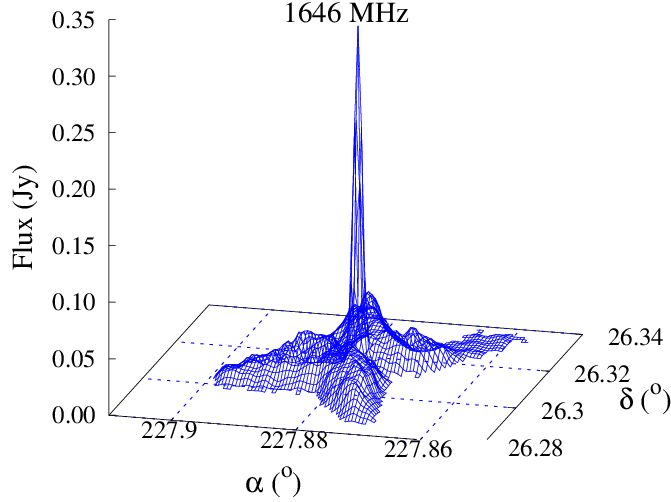}
\caption{The same as in Fig. \ref{fig01}, but for 1646 MHz.}
\label{fig02}
\end{figure*}

\begin{figure*}[ht!]
\centering
\includegraphics[width=0.48\textwidth]{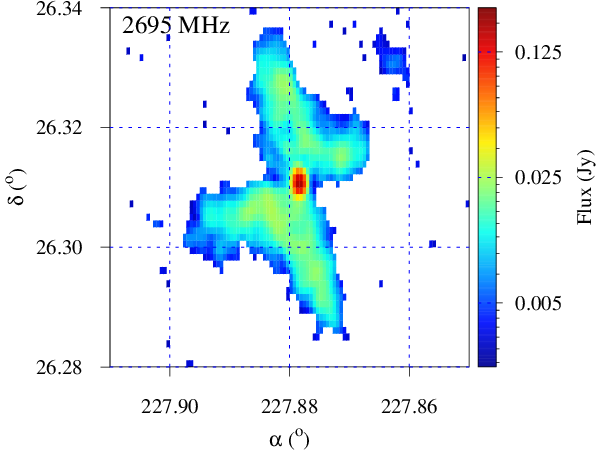}
\hfill
\includegraphics[width=0.46\textwidth]{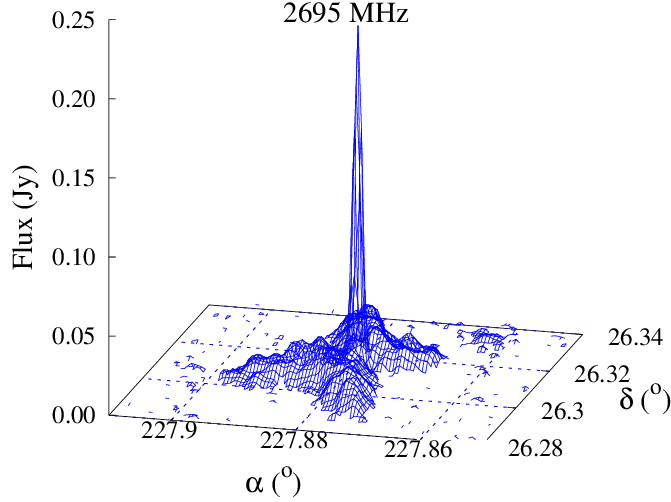}
\caption{The same as in Fig. \ref{fig01}, but for 2695 MHz.}
\label{fig03}
\end{figure*}

From the available online-data, we used Flexible Image Transport System (FITS) data files containing the flux densities in Jy (1 Jy = $10^{–26} W m^{–2} Hz^{–1}$) of some radio source. We described the structure of FITS format, as well as what is useful for our investigation, in our previous papers (Arsenić et al. 2022; Borka Jovanović et al. 2023a,2023b,2023c). The observed data are available in An Atlas of DRAGNs i.e. the ''3CRR'' sample of Laing, Riley \& Longair (1983). This sample contains not only the FITS files, but it also gives the readers very useful information from the literature on the DRAGNs, with tables and references. There are the Introductory Pages, Description Pages (with full details) and the Listings of individual DRAGNs. Also, we used astronomical database compiled by NASA and IPAC with the information and bibliographic references regarding astronomical objects. Therse data are provided at: (1) J. P. Leahy, A. H. Bridle \& R. G. Strom, An Atlas of DRAGNs (2013): \url{http://www.jb.man.ac.uk/atlas/} (Leahy, Bridle and Strom 2013), and (2) NASA/IPAC Extragalactic Database: \url{http://ned.ipac.caltech.edu/} (Mazzarella and the NED Team 2002).

\clearpage

We used observations of 3C 315 at three different frequencies, 1417 MHz, 1646 MHz and 2695 MHz (Alexander and Leahy 1987). The calculation method, that we developed, was first published in Borka (2007), and further elaborated in Borka Jovanović (2012) and in Borka Jovanović et al. (2012).

From the observations of 3C 315 at the three frequencies, 1417 MHz (21.2 cm), 1646 MHz (18.2 cm) and 2695 MHz (11.1 cm), we determined the contours which represent the lower boundaries of the source. We found that the minimal fluxes are the following: $S_{\nu,min}$ = 0.0018 Jy at 1417 MHz, $S_{\nu,min}$ = 0.0020 Jy at 1646 MHz and $S_{\nu,min}$ = 0.0022 Jy at 2695 MHz. The areas of 3C 315, with the flux density distributions (in Jy) over these areas, we presented by two-dimensional and three-dimensional plots in Figs. \ref{fig01} - \ref{fig03}.

\section{SPECTRAL INDEX DISTRIBUTION}

The expression which represents the flux density $S_\nu$ as a function of frequency $\nu$ is given by the expression:

\begin{equation}
S_\nu \sim \nu^{-\alpha},
\label{equ01}
\end{equation}   

\noindent where $\alpha$ is a constant, called the 'radio spectral index'.

The radio spectral index $\alpha$ can be obtained using the flux density at different frequencies and taking the negative slope of the relation (\ref{equ01}). So we calculate it by the following equation:

\begin{equation}
\alpha = - \dfrac{\log\left({\dfrac{S_{\nu_1}}{S_{\nu_2}}}\right)}{\log\left({\dfrac{\nu_1}{\nu_2}}\right)}.
\label{equ02}
\end{equation}

The spectral index is used for classification of radio sources and studying the origin of radio emission:

\begin{itemize}[label=-,nosep]
\item if $\alpha > 0.1$ the emission is non-thermal (synchrotron) and it means that it does not depend on the temperature of the source, 
\item for $\alpha < 0$ it is thermal and depends only on the temperature of the source.
\end{itemize}

\begin{figure*}[ht!]
\centering
\includegraphics[width=0.49\textwidth]{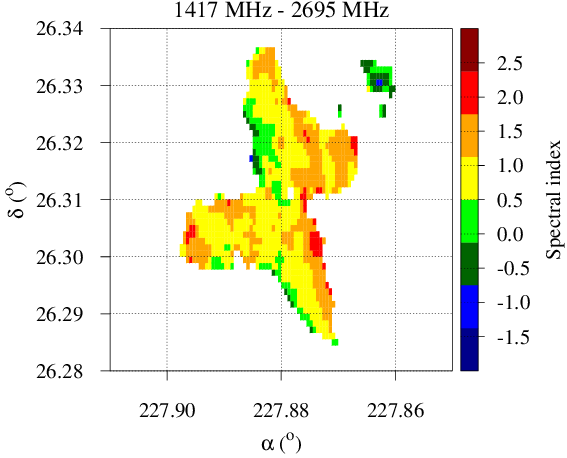}
\hfill
\includegraphics[width=0.49\textwidth]{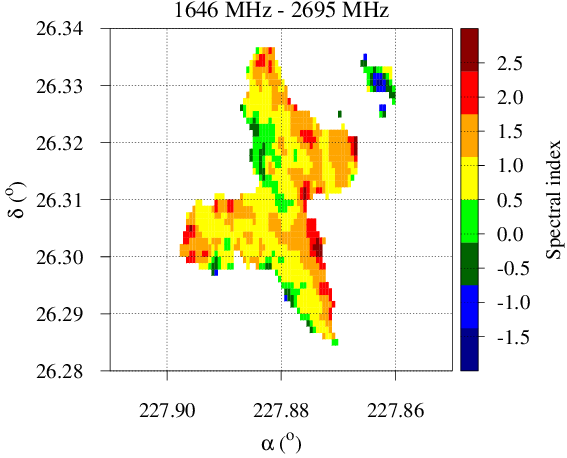}
\caption{Spectral indices between 1417 and 2695 MHz (left) and between 1646 and 2695 MHz (right) over the area of 3C 315.}
\label{fig04}
\end{figure*}

\begin{figure*}[ht!]
\centering
\includegraphics[width=0.95\textwidth]{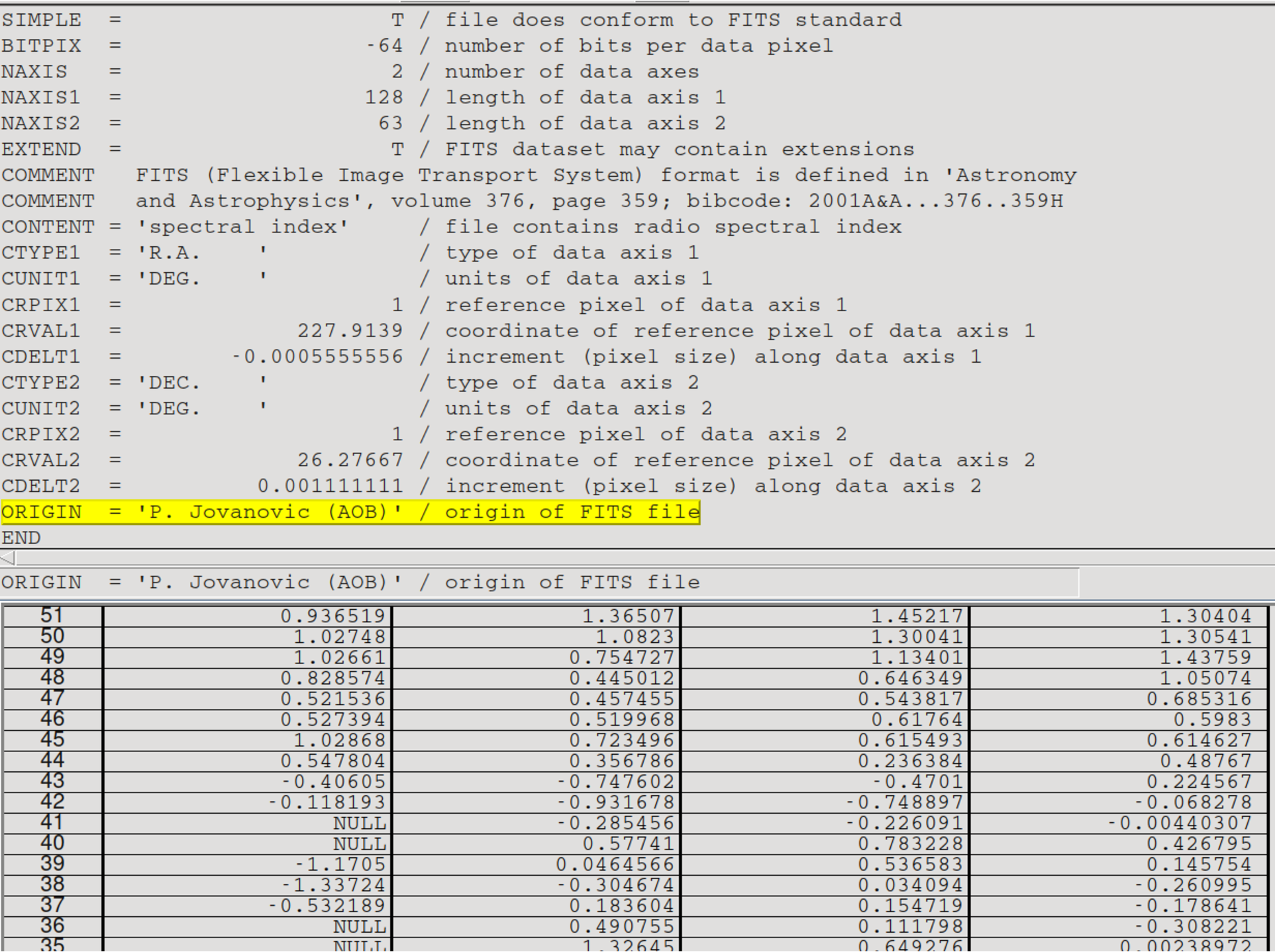}
\caption{Header unit (upper part) and a part of data unit (lower part) of the FITS data that we have obtained from our calculated spectral indices over the area of 3C 315.}
\label{fig05}
\end{figure*}

We calculated spectral indices 1417 MHz - 2695 MHz, as well as 1646 MHz - 2695 MHz, over the whole area of 3C 315, and we show how they change over this area in Fig. \ref{fig04}. As expected, these two maps are very similar, because both frequencies 1417 and 1646 MHz (which we combined with 2695 MHz) are very near within the frequency range.

While reading the values of radio spectral index $\alpha$ from the colorbar in Fig. \ref{fig04}, it is noticeable that over the area of 3C 315 it takes values from negative to positive, meaning this: $\alpha > 0$ corresponds to non-thermal mechanism of radiation while $\alpha < 0$ corresponds to thermal mechanism of radiation. For the spectral index of zero value, the flux density is independent of frequency, and the spectrum is said to be flat.

The numerical values of the spectral indices that we have obtained, we further use as a new data. We present it in Fig. \ref{fig04}. The header unit, as well as the part of data unit (table) of this FITS data we show in Fig. \ref{fig05}.

\begin{figure*}[ht!]
\centering
\includegraphics[width=0.65\textwidth]{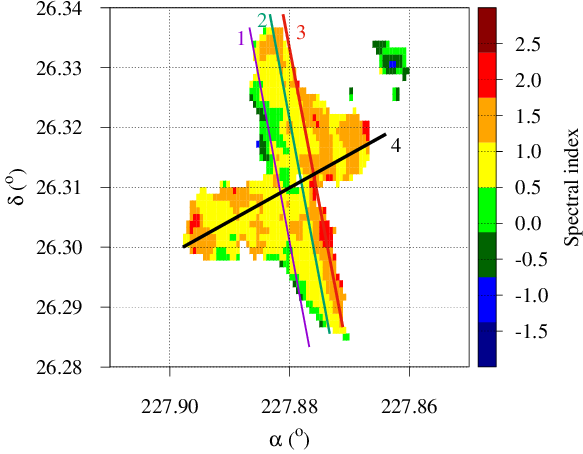} \\
\vspace{0.6cm}
\includegraphics[width=0.48\textwidth]{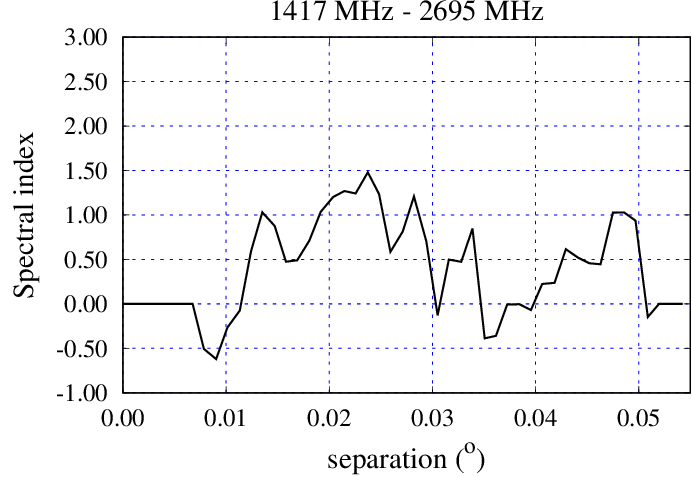}
\hfill
\vspace{0.5cm}
\includegraphics[width=0.48\textwidth]{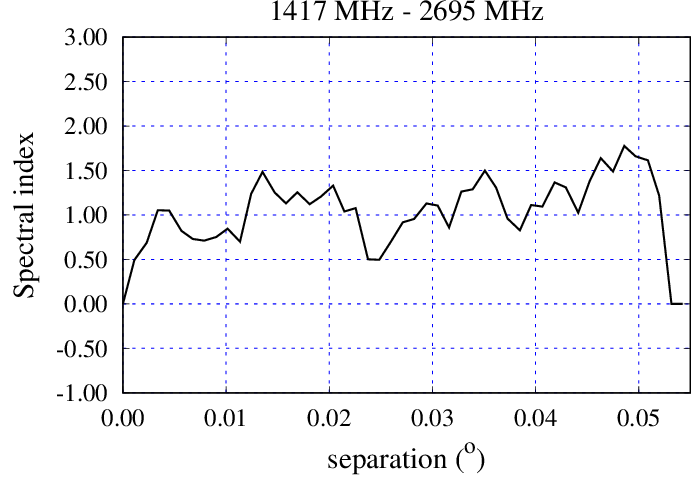}
\includegraphics[width=0.48\textwidth]{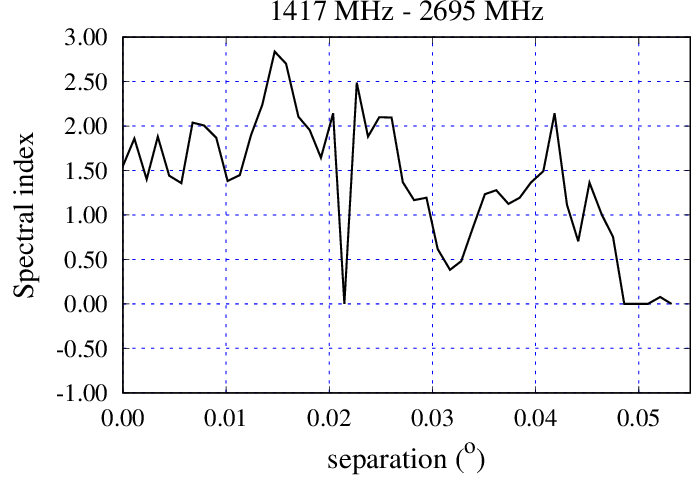}
\hfill
\includegraphics[width=0.48\textwidth]{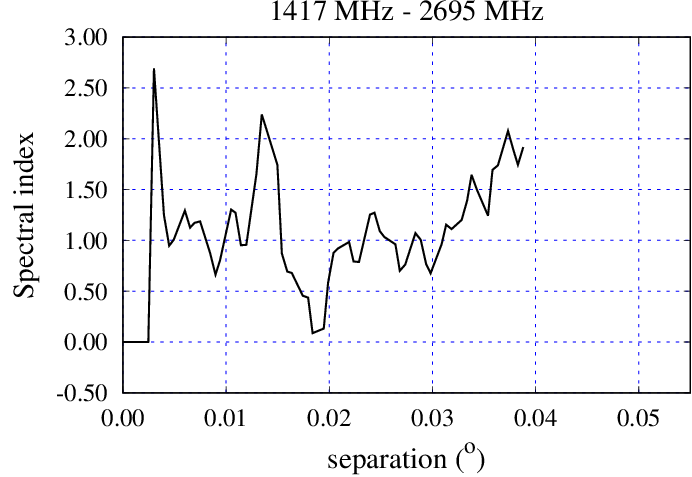}
\caption{Spectral index map of 3C 315 between 1417 and 2695 MHz with 4 designated directions: thin purple line, thin green line, thick red line and thick black line. Below map, the four spectral index profiles which correspond to these directions are given.}
\label{fig06}
\end{figure*}

\clearpage

With aim to investigate the distribution of $\alpha$ along some specific directions, we draw the profile lines over the spectral index map 1417 MHz - 2695 MHz (see Fig. \ref{fig06}). The chosen directions, three NNE-SSW and one SE-NW, are the following:

\begin{enumerate}[nosep]
\item $(\alpha, \delta)_{start} = (227^\circ.8767, 26^\circ.28334)$,\\
$(\alpha, \delta)_{end} = (227^\circ.8867, 26^\circ.33667)$;
\item $(\alpha, \delta)_{start} = (227^\circ.8733, 26^\circ.28556)$,\\
$(\alpha, \delta)_{end} = (227^\circ.8833, 26^\circ.33889)$;
\item $(\alpha, \delta)_{start} = (227^\circ.8711, 26^\circ.28667)$,\\
$(\alpha, \delta)_{end} = (227^\circ.881, 26^\circ.33889)$;
\item $(\alpha, \delta)_{start} = (227^\circ.8639, 26^\circ.31889)$,\\
$(\alpha, \delta)_{end} = (227^\circ.8978, 26^\circ.3)$
\end{enumerate}

We can notice that 3C 315 has two pairs of lobs. The direction of the main lobes is NNE-SSW, and that is why we show three profiles along just this direction. Although the spectral index has strong variability over all three studied paths along this direction, nevertheless it can be seen that its average value is different along each of these paths, and that it increases from the first path towards the third one. The first path is taken along valley of spectral indices map, and average value of spectral index along this direction is about 0.5. The second path is parallel to the first one, chosen to connect north-west wormspot and AGN, and has average value of $\alpha$ about 1.1. The third path, also parallel to the first one, and taken along outer border of spectral index map, has average value of $\alpha$ along this direction about 1.5. Therefore, the regions around this third path are with the steepest spectrum along the NNE-SSW direction, meaning that this edge is more recent.

The forth line is taken along SE-NW axis. The lobes on this axis are not so well defined and are more curved, known as faint extensions of the main lobes i.e. wings. Average value of spectral index along this direction is about 1.1 which is similar to the middle path along the NNE-SSW direction, but variation of spectral indices is much larger than in the previous three cases.

As is it can be seen from all four presented profiles, the spectral index is almost always higher than zero, except in only few small parts where it is negative. This indicates that the non-thermal (synchrotron) emission is by far the most dominant radiation  mechanism over the whole source.

Also, 3C 315 contains pronounced and bright radio core (see Figs. \ref{fig01} - \ref{fig03}), but only a weak hotspots. The outer structure is relaxed.

\section{DISCUSSION AND CONCLUSIONS}

We used the available flux densities of 3C 315 at 1417, 1646 and 2695 MHz. We provide the spectral index distribution derived between two frequencies: 1417 and 2695 MHz, as well as between 1646 and 2695 MHz. At all three frequencies the flux structure is characterized with no obvious hotspots and the core dominates the flux density distribution. It is noticeable from the spectral index map that this tendency is even more pronounced. We clearly show here that the synchrotron radiation is the dominant emission mechanism over the majority of the area of this radio galaxy.

Some authors propose that X-shaped radio sources may be the result of a few possibilities (Leahy and Williams 1984; Capetti 2002; Merritt and Ekers 2002; Gopal-Krishna, Biermann and Wiita 2003):
\begin{enumerate}[label=\arabic*),nosep]
\item the AGN has undergone two separate epochs of activity (the brighter lobes define the axis of current activity),
\item recent collision or merger between two supermassive black holes, which can produce the extra set of jets and lobes;
\item the main jets expanded into an asymmetric medium, and in that way generating an additional pair of radio lobes.
\end{enumerate}

We are led to the conclusion that results of this study will be helpful for understanding the evolutionary process of the 3C 315 radio source. Due to the fact that the host galaxy of 3C 315 (Koff et al. 2000) is accompanied by an elliptical galaxy and both of which are located inside a cluster (an anisotropy in the gas environment is related to the ellipticity of the host galaxy), probably the environment plays the most significant role in the evolution of this object.

\paragraph{Acknowledgments.} This work is supported by Ministry of Science, Technological Development and Innovations of the Republic of Serbia through the Project contracts No. 451-03-66/2024-03/200017 and 451-03-66/2024-03/200002.

\vskip-.5cm

\references

Alexander, P., Leahy, J. P.: 1987, \journal{Mon. Not. R. Astron. Soc.}, \vol{225}, 1.

Arsenić, A., Borka, D., Jovanović, P., Borka Jovanović, V.: 2022, \journal{Publ. Astron. Obs. Belgrade}, \vol{102}, 265.

Borka, V.: 2007, \journal{Mon. Not. R. Astron. Soc.}, \vol{376}, 634.

Borka Jovanović, V.: 2012, \journal{Publ. Astron. Obs. Belgrade}, \vol{91}, 121.

Borka Jovanović, V.: Borka, D., Skeoch, R., Jovanović, P., 2012, \journal{Publ. Astron. Obs. Belgrade}, \vol{91}, 255.

Borka Jovanović, V., Borka, D., Arsenić, A., Jovanović, P.: 2023a, \journal{Adv. Space. Res.}, \vol{71}, 1227.

Borka Jovanović, V., Borka, D., Jovanović, P.: 2023b, \journal{Contrib. Astron. Obs. Skalnate Pleso}, \vol{53}, 188.

Borka Jovanović, V., Borka, D., Jovanović, P.: 2023c, \journal{PoS}, \vol{BPU11}, 043.

Capetti, A. et. al.: 2002, \journal{Astron. Astrophys.}, \vol{394}, 39.

de Koff, S. et al.: 2000, \journal{Astrophys. J. Suppl. Series}, \vol{129}, 33.

Gopal-Krishna, Biermann, P. L., Wiita, P. J.: 2003, \journal{Astrophys. J.}, \vol{594}, L103.

Hogbom, J. A.: 1979, \journal{Astron. Astrophys. Suppl.}, \vol{36}, 173.

Laing, R. A., Riley, J. M., Longair, M. S.: 1983, \journal{Mon. Not. R. Astron. Soc.}, \vol{204}, 151.

Leahy, J. P.: 1993, ''DRAGNs'', in Proceedings: Jets in Extragalactic Radio Sources, Eds. Roser, H.-J. and Meisenheimer, K., Lecture Notes in Physics, \vol{421}, 1.

Leahy, J. P., Williams, A. G.: 1984, \journal{Mon. Not. R. Astron. Soc.}, \vol{210}, 929.

Leahy, J. P., Pooley, G. G., Riley, J. M.: 1986, \journal{Mon. Not. R. Astron. Soc.}, \vol{222}, 753.

Leahy, J. P., Bridle, A. H., Strom, R. G.: 2013, An Atlas of DRAGNs - online: \url{http://www.jb.man.ac.uk/atlas/}.

Marecki, A.: 2012, \journal{PoS}, \vol{RTS2012}, 24.

Mazzarella, J. M. and the NED Team: 2002, ''Using the NASA/IPAC Extragalactic Database (NED) and Federated Virtual Observatory Archives for Multiwavelength Studies of AGN'', in Proceedings: AGN Surveys, Eds. Green, R. F., Khachikian, E. Ye. and Sanders, D. B., \journal{Astron. Soc. Pacific Conference Series}, \vol{284}, 379.

Merritt, D., Ekers, R. D.: 2002, \journal{Science}, \vol{297}, 1310.

Northover, K. J. E.: 1976, \journal{Mon. Not. R. Astron. Soc.}, \vol{177}, 307.

Saripalli, L., Subrahmanyan, R., Laskar, T., Koekemoer, A.: 2007, \journal{PoS}, \vol{MRU}, 130.

Saripalli, L., Subrahmanyan, R.: 2009, \journal{Astrophys. J.} \vol{695}, 156.

Yang, X., Joshi, R., Gopal-Krishna, An, T., Ho, L. C., Wiita, P. J., Liu, X., Yang, J., Wang, R., Wu, X.-B., Yang, X.: 2019, \journal{Astrophys. J. Suppl.}, \vol{245}, 17.

\endreferences

\end{document}